## Title

# Peculiarities and evolution of Raman spectra of multilayer Ge/Si(001) heterostructures containing arrays of low-temperature MBE-grown Ge quantum dots of different size and number density: Experimental studies and numerical simulations

## Authors

Mikhail S. Storozhevykh[1], Larisa V. Arapkina[1], Sergey M. Novikov[2], Valentyn S. Volkov[2], Aleksey V. Arsenin[2], Oleg V. Uvarov[1], Vladimir A. Yuryev[1]

## Affiliations

[1]A. M. Prokhorov General Physics Institute of the Russian Academy of Sciences, 38 Vavilov Street, Moscow, 119991, Russia
[2]Center for Photonics and 2D Materials, Moscow Institute of Physics and Technology, 9 Institutskiy per., Dolgoprudny, Moscow Region, 141707, Russia

E-mail: storozhevykh@kapella.gpi.ru

## Abstract

Ge/Si(001) multilayer heterostructures containing arrays of low-temperature self-assembled Ge quantum dots and very thin $Si_xGe_{1-x}$ layers of varying composition and complex geometry have been studied using Raman spectroscopy and scanning tunneling microscopy. The dependence of Raman spectra on the effective thickness of deposited Ge layers has been investigated in detail in the range from 4 to 18 Å. The position and shape of both Ge and SiGe vibrational modes are of great interest since they are closely related to the strain and composition of the material that plays a role of active component in perspective optoelectronic devices based on such structures. In this work, we present an explanation for some peculiar features of Raman spectra, which makes it possible to control the quality of the grown heterostructures more effectively. A dramatic increase of intensity of both the Ge–Ge and Si–Ge bands for the structure containing Ge layers of 10 Å and anomalous shift and broadening of the Si–Ge band for structures comprising Ge layers of 8 and 9Å thick were observed. In our model, the anomalous behavior of the Raman spectra with the change of thickness of deposited Ge is connected with the flatness of Ge layers as well as transitional SiGe domains formed via the stress-induced diffusion from {105} facets of quantum dots. The conclusions are supported by the STM studies and the numerical calculations.

## 1 | INTRODUCTION

Semiconductor heterostructures and superlattices of different compounds entrench themselves in many fields dealing with electronic and optoelectronic devices.[1–8] Some of them include very thin layers down to several angstroms thick. Controlling the quality of such structures is an extremely complex and important task because tiny discrepancies in composition, strains or morphology can lead to disastrous changes in electrical and optical properties. The Si/Ge(001) based structures studied in this work attract special attention due to the possibility to integrate the process of their fabrication into the standard CMOS technology. Moreover, the feasibility of such structures as light emitters, detectors and lasers was shown in many works.[5,6,9–11]

Raman spectroscopy is a favorite method of analysis of the internal structure of solid materials, i.e. their crystallinity, composition, stress, defects, etc., since it is a fast, cheap and non-destructive technique imposing no special requirements on processing of a sample.[12–14] However, the interpretation of features of Raman spectra often requires additional knowledge about the sample obtained using other experimental methods or derived from theoretical calculations. More particularly, the well-known dependence of Raman spectra on stress and composition of a material can be significantly disturbed by the unique morphology of thin layers or inclusions under investigation. Some features of Raman spectra caused by the complex geometry of layers in a heterostructure, e.g., a non-linear dependence of the intensity of the Raman peak assigned to the Si–Ge mode on the absolute amount of $Si_xGe_{1-x}$ solution observed in our experiment, may arise in any structures having thin non-flat layers of a mixed composition and these effects can lead to wrong conclusions about structural properties of the studied sample. It can be critical when the change of a layer thickness by 10 % leads to the increase of Raman response at the corresponding wave number by 1000 %.

Ge/Si heterostructures with arrays of self-assembled Ge quantum dots (QDs) studied in this work are solid-state structures with very thin layers of variable composition, which at the same time are not flat and have a complex geometry. Low-temperature QDs used in our experiment represent hut-shaped clusters of 6 to 12 Å in height faceted by {105} planes having a pure germanium core surrounded with a shell of mixed composition and situated on the patched germanium wetting layer of 4 to nearly 6 Å in thickness. The interpretation of Raman spectra of such structures is a substantive challenge bound up with the study of the morphology and growth process of layers with QDs. We have to refer to the process of intermixing between Si and Ge during epitaxial growth[15–19] to explain peculiar features of Raman spectra evolution with the increase of the effective thickness of deposited Ge layers ($h_{Ge}$). Stress distribution in Ge/Si heterostructures with QDs is also of great interest because it leads to both positive and negative effects. The former is vertically correlated growth of QDs in multilayer structures,[20–26] while the latter is the formation of defects, which can drastically impair the optical and electrical characteristics of a sample.[27]

In this research, we attempt to associate the morphology of Ge QDs and transitional SiGe layers[28] with not only vivid but also barely discernible features of Raman spectra. Experimental results are supported by numerical calculations to clarify some disputable issues as far as possible. It is worth noting that all obtained results are closely related to the character of growth of Ge QDs at particular conditions. However, all conclusions and used approaches can be expanded to other growth conditions or materials with success.

## 2 | METHODS AND EQUIPMENT

The samples used in the experiment were grown by MBE in an EVA32 chamber of Riber SSC 2 surface-science center at ultra-high vacuum (UHV) conditions. Ge and Si sources with electron-beam evaporation were used for layers deposition on p-type (ρ = 12 Ω·cm) Si (100) substrates. The substrates were subjected to preliminary chemical treatments followed by annealing at the temperature 600 °C for removing adsorbates from the silicon surface lasted for 6 hours in the high-vacuum preliminary cleaning chamber. The final deoxidation lasted for 30 minutes was carried out in the growth chamber under a flux of Si atoms of ≤ 0.1 Å/s at 800 °C. After that, a Si buffer layer of 100 nm in thickness was grown on the prepared surface at the substrate temperature of 650 °C. Layers of Ge QDs were grown at the temperature of 360 °C, while Si spacers of 50 nm in thickness and cap layers were grown at the temperature of 530 °C.[28,29]

The Ge coverage $h_{Ge}$ was varied from 4 to 18 Å for different samples; the deposition rate and $h_{Ge}$ were controlled using an Inficon XTC 751-001-G1 film thickness monitor (Leybold-Heraeus) equipped with the graduated in-advance quartz sensor installed in the MBE chamber.[29,31] Growth temperature was monitored using IMPAC IS12-Si IR pyrometer (LumaSense Technologies) and *in-situ* graduated tungsten-rhenium thermocouples mounted in a vacuum near the rear side of samples.[1]

Every sample had a single-layer uncapped twin studied using an STM GPI-300 UHV scanning tunneling microscope (Sigma Scan) linked with the MBE chamber via a UHV transfer line.

TEM images of the capped and multilayer structures were obtained using a Libra-200 FE HR instrument (Carl Zeiss) after the ion treatment in argon plasma using a Model 1010 ion mill (E. A. Fischione Instruments) for obtaining thin defectless lamellae.

STM and TEM images were processed using the WSxM software.[32]

A confocal scanning a Horiba LabRAM HR Evolution Raman microscope with linearly polarized excitation at the wavelength of 632.8 nm (spot size is ~ 0.43 μm), 1800 gr/mm diffraction grating (spectral resolution less than 1 $cm^{-1}$), and ×100 objective (N.A. = 0.90) was used for Raman spectra measurements. We used unpolarized detection in order to have a significant signal-to-noise ratio. The Raman spectra were recorded with 1.8, 0.7 and 0.35 mW incident powers and an integration time of 10 s at each point. The statistics were collected from at least 10 points from different parts of the sample.

The atomistic model of thin layers and nanoparticles of SiGe containing not Si–Si or Ge–Ge but only Si–Ge bounds embedded in Si matrix was used in numerical simulations. The supercell of 72 atoms was chosen as the best balanced between time consumption and precision of computation. Geometry optimization of the unit cell and Raman spectra calculation were carried out using CASTEP module of BIOVIA Materials Studio software. Both LDA and GGA functionals were applied with ultra-fine quality. Raman intensities were calculated via Hessian basing on perturbations of electric field.

## 3 | RESULTS AND DISCUSSION

### 3.1 | Experiment

All the samples used in the experiment are heterostructures with 5 layers of Ge divided by 50 nm of Si (Fig. 1). Each Ge layer represents hut-shaped Ge QDs situated on a patched Ge wetting layer of ~ 5 Å in thickness. The total effective thickness of deposited Ge ($h_{Ge}$) for every layer varied from 4 to 18 Å is the only distinctive feature of the samples.

[1] For sample preparation details and process parameters, see, e.g., Refs. 29,31,35.

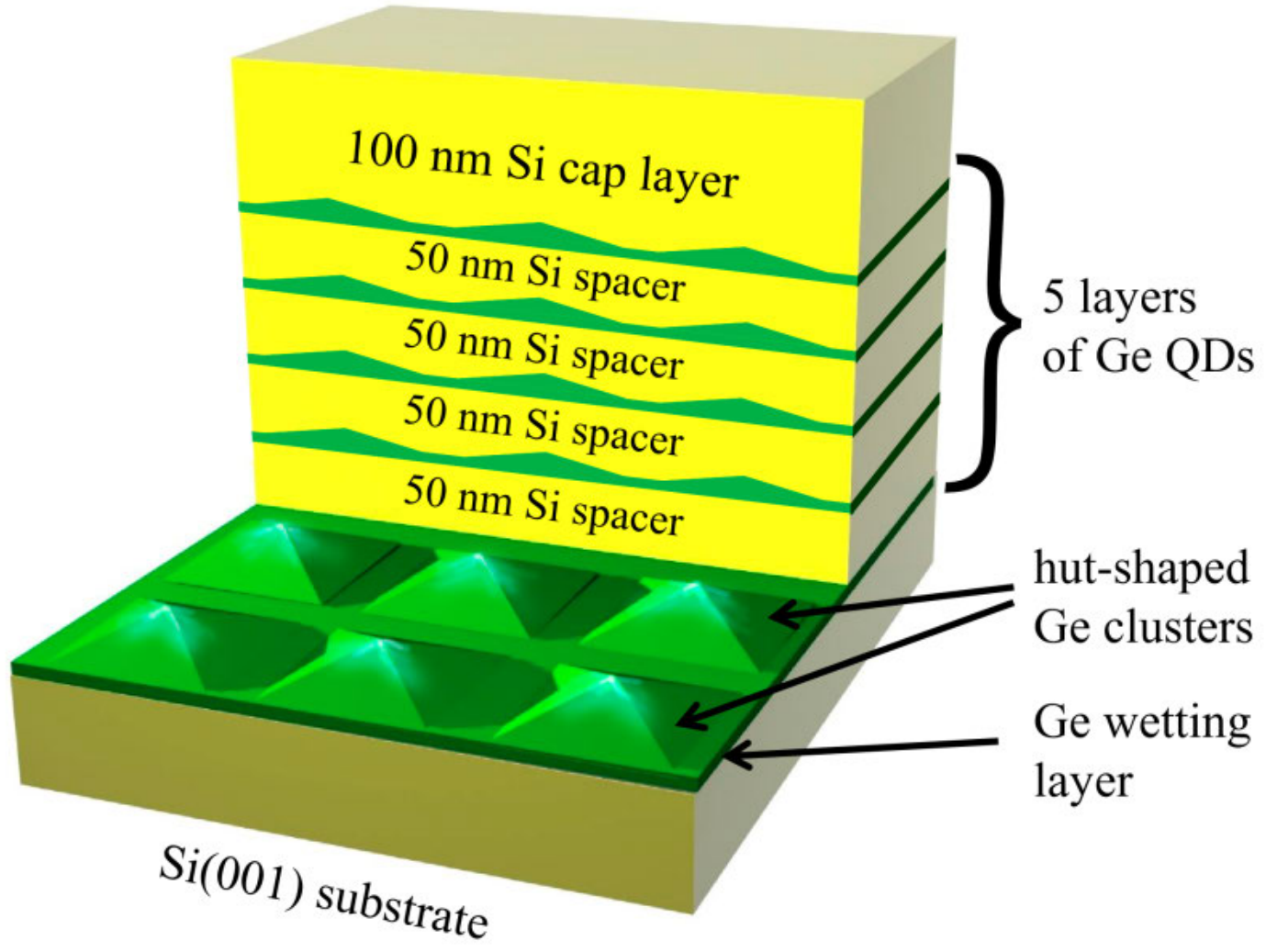

**FIGURE 1** Generalized schematic view demonstrating the cross-section of multilayer Ge/Si(001) heterostructures used in the experiment; the first layer of Ge QDs is exposed to show the typical morphology of Ge layers containing hut-shaped clusters situated on the wetting layer.

The evolution of QDs with the increasing $h_{Ge}$ is shown in Fig. 2. STM study of single-layer uncapped samples allowed us to reveal break points subdividing the growth of Ge QDs into several stages.[29–31,33,34] Although Ge clusters nucleate on the wetting layer at $h_{Ge}$ = 5.1 Å and the first tiny QDs form at $h_{Ge}$ = 5.5 Å,[34-36] in fact the first stage of the Ge QD array growth starts at $h_{Ge} \approx$ 6 Å when it forms as an ensemble of 3D clusters.[36] At the second stage, which starts at $h_{Ge} \approx$ 8 Å, QDs mainly cover the wetting layer area and a QD array reaches the maximum number density of Ge huts.[36] The third stage starts at $h_{Ge} \approx$ 10 Å when almost all QDs closely contact with their neighbors that corresponds to the beginning of QDs coalescing in an array.[33,34,36] Then, at $h_{Ge} \approx$ 14 to 15 Å, QDs completely coalesce,[33,34,36] although rare domains of the wetting layer are still can be found,[31,34] in which new huts go on nucleating.[31] Finally, a virtually continuous nanocrystalline film forms at $h_{Ge} \sim$ 18 Å,[33,34,36] which however contains a few deep pits with wetting layer patches seen on their bottoms.[34] The Raman measurements were performed for each stage listed above.

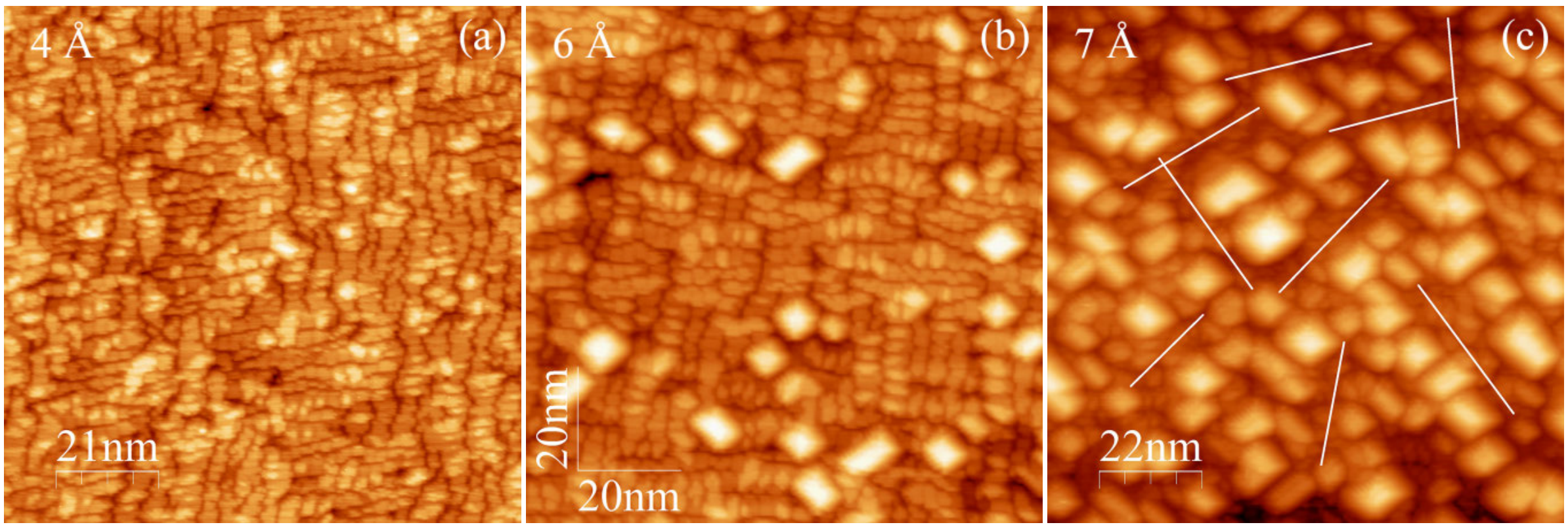

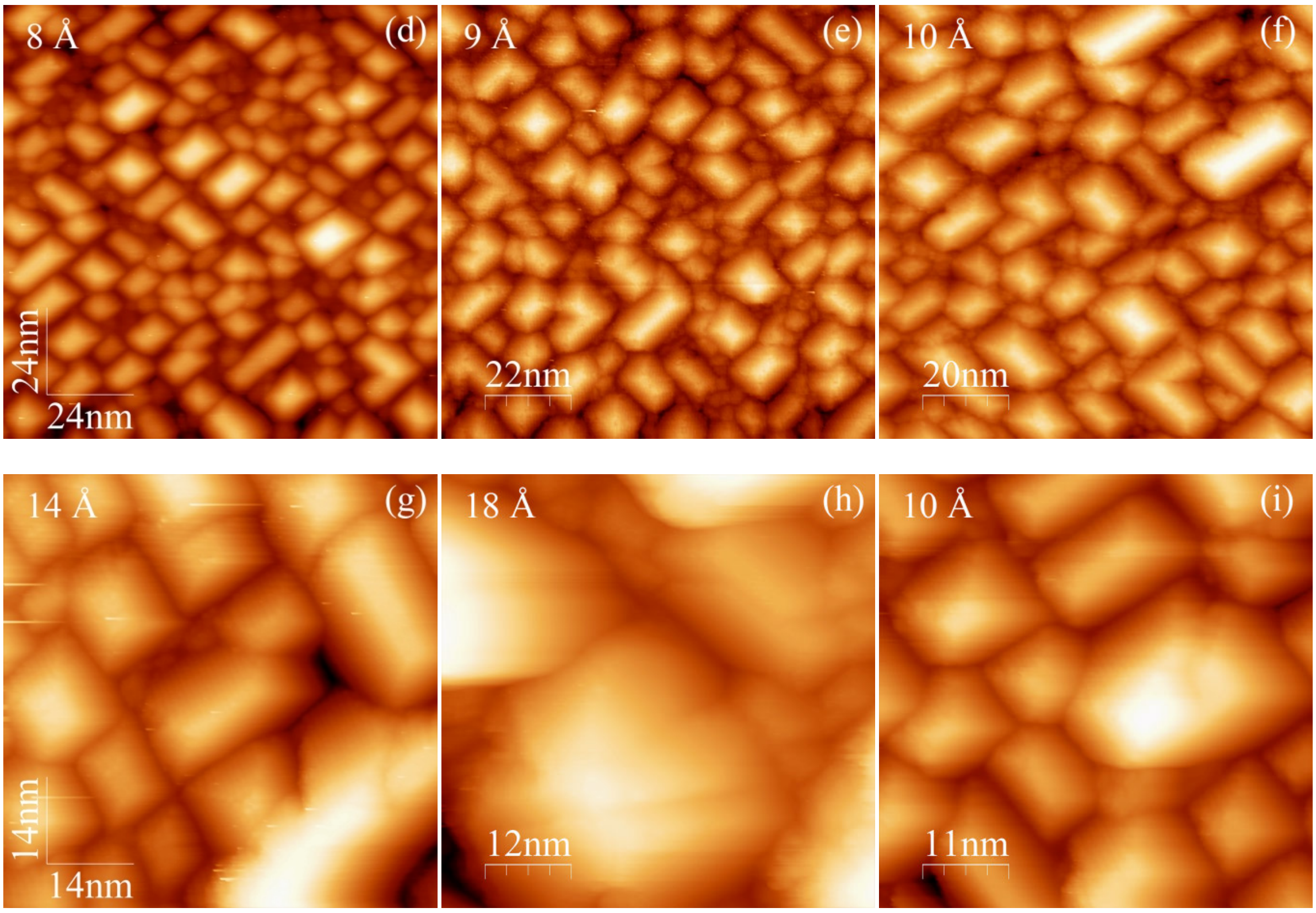

**FIGURE 2** STM images of Si (001) surface covered by Ge of different thickness at 360 °C; $h_{Ge}$ is specified at upper left corner of every image; white lines at the image for $h_{Ge}$ = 7Å demonstrate extended domains of wetting layer.

Survey Raman spectra demonstrating both Ge–Ge and Si–Ge vibrational modes for the samples under investigation are presented in Fig. 3. The strong Si–Si peak at 520 cm$^{-1}$ that is not shown in the plot remains unchanged in our measurements. We have subtracted base lines and carried out the decomposition by peaks to exclude the contribution of Si 2TA phonon and to reveal the true positions of the Ge–Ge vibrational mode in the spectra. Note that the Si 2TA line cannot be directly subtracted using the reference sample because the presence of stressed Si in the spacers between Ge layers slightly distorts the Si 2TA line. The corresponding spectra for all samples are shown in Fig. 4. The position of the peak assigned to the Ge–Ge mode is closely related to the stress in Ge domains in a sample. For unstrained Ge, it is situated at ~ 300 cm$^{-1}$ and it undergoes red shift at a tensile strain and blue shift at a compressive strain.

The possible impact of a sample heating during the experiment was analyzed by carrying out measurements at different incident power. Raman spectra of the Ge reference sample and the SiGe structure with quantum dots obtained at different incident power are presented in Figure S2. The results shows that the impact of a sample heating on the position of both Ge–Ge and Si–Ge peak is negligible in our experiments.

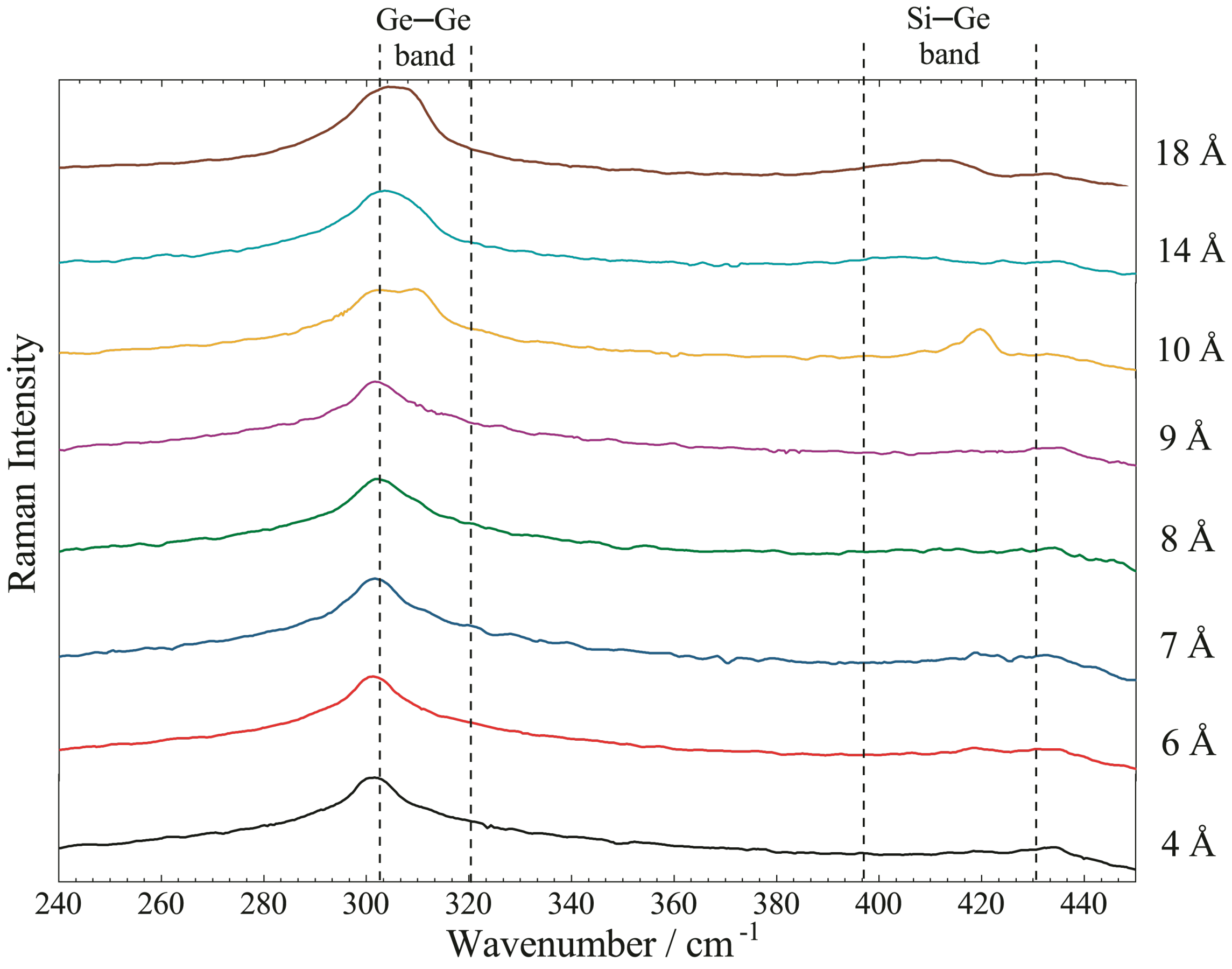

**FIGURE 3** Survey Raman spectra representing Ge–Ge and Si–Ge vibrational modes for all the studied samples; the borders of Ge–Ge and Si–Ge band ranges are depicted by dashed lines.

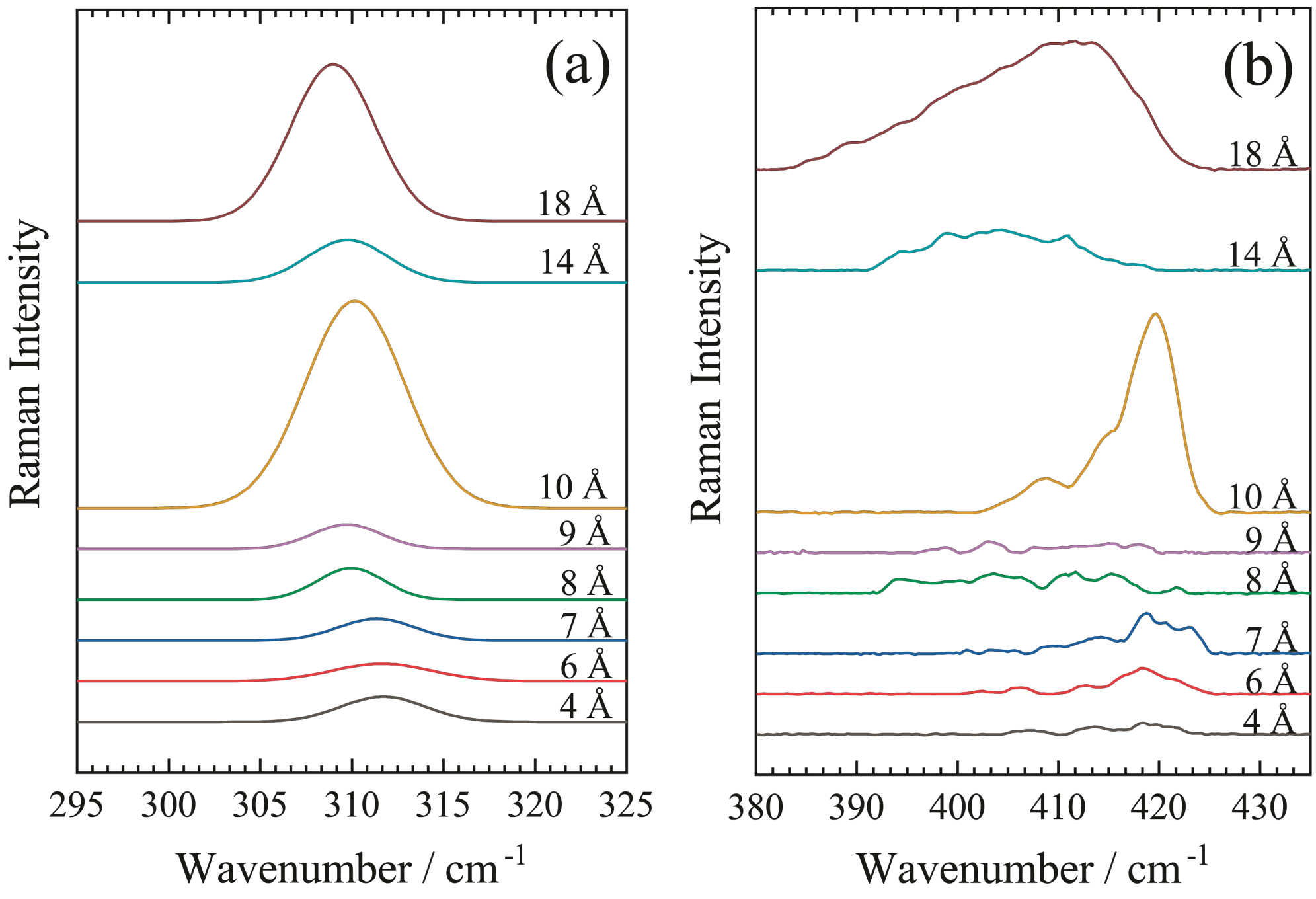

**FIGURE 4** Raman spectral bands related to Ge–Ge (a) and Si–Ge (b) vibrational modes.

### 3.2 | Ge–Ge band analysis

We start from $h_{Ge}$ = 4 Å that is close to the thickest wetting layer still having no nanoclusters on its surface (Fig. 2).[35,36]

Although the topmost atomic layer of uncapped Ge wetting layer patches was found to be relaxed,[33,35-37] the wetting layer buried in the multilayer structure is significantly stressed and has a lattice parameter close to that of Si.[28] The Ge–Ge mode in the wetting layer has the largest blue shift among all of the samples (Fig. 4a, $h_{Ge}$ = 4 Å) and continuously decreases with the increase of the thickness of deposited Ge since the start of forming of 3D Ge clusters gives rise to the gradual relaxation of stress in the buried Ge layer. The least blue shift of the Ge–Ge mode in the sample with $h_{Ge}$ = 18 Å corresponds to the least stressed Ge among all the samples in this study.

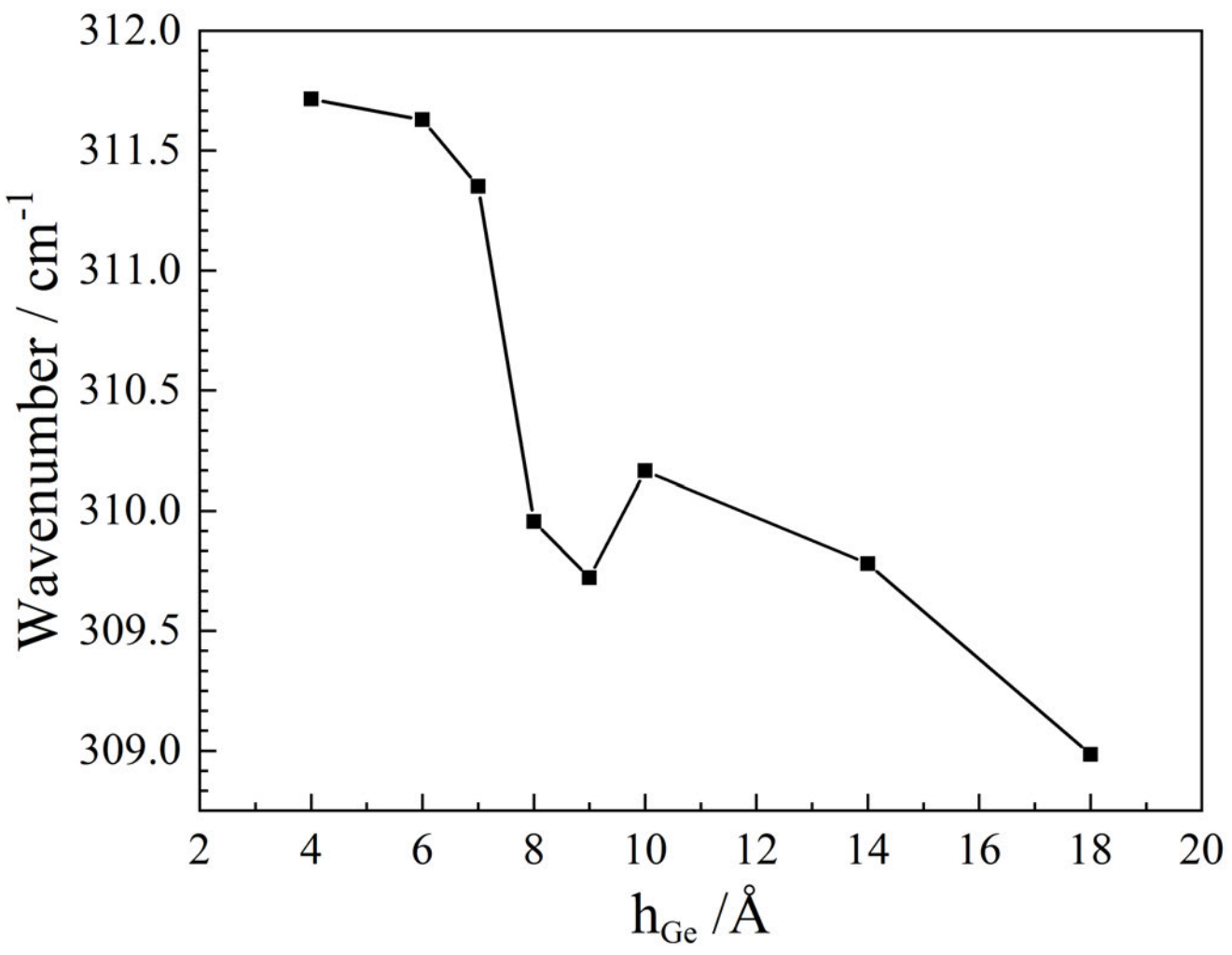

**FIGURE 5** The dependence on the position of the Ge–Ge vibrational band on $h_{Ge}$.

**TABLE 1** Dependence of the lateral compression of Ge lattice on the thickness of Ge layers.

| $h_{Ge}$, Å | 4 | 6 | 7 | 8 | 9 | 10 | 14 | 18 |
|---|---|---|---|---|---|---|---|---|
| Compression, % | 1.53 | 1.51 | 1.44 | 1.15 | 1.12 | 1.21 | 1.14 | 0.97 |

We have carried out the simulation of Raman spectra for bulk Ge crystal under the laterally applied pressure to be able to derive numerical values of the stress of Ge in the structures under investigation. The obtained results and parameters of the simulation are presented in Fig. S1 and Table S1, respectively. The estimation of the stress appeared in Ge layers in the experimental samples is given in Table 1 as a percentage of the lattice parameter change in comparison with the lattice parameter of the unstressed Ge lattice. The results show that Ge relaxes significantly even in the structure having only wetting layers taking into account the lattice mismatch between Si and Ge equal to 4.2%.

The stress in the Ge layer changes non-linearly in relation to its thickness. It decreases slowly at low values of $h_{Ge}$ (4 to 7 Å) (Fig. 5) and in the $h_{Ge}$ range from 8 to 18 Å, but it rapidly drops down for $h_{Ge}$ from 7 to 8 Å. We suppose it is due to the change in the number density of QDs and the distance between them. There are quite broad regions of highly stressed virgin wetting layer at $h_{Ge}$ = 4 to 7 Å, which obviously contributes to the signal, while at $h_{Ge}$ = 8 Å, the surface is mainly covered by QDs (Fig. 2). Long valleys with patched wetting layer on bottoms are shown by white lines in the STM image of the Ge layer at $h_{Ge}$ = 7 Å. However, such lines cannot be drawn in the STM image of Ge layer at $h_{Ge}$ = 8 Å. QDs nucleating on relaxed tops of wetting layer patches,[35] and then growing to prevent stress occurrence in the layer,[37] reconstruct the wetting layer by covering patches and trenches by their bases.[36,37] Thus, when overgrown by a Si layer, they protect the areas of the Ge layer from the strain induced by the growing Si film. Buried Ge QDs are less stressed in comparison with the wetting layer due to a higher thickness and intermixing of Si and Ge from {105} facets of them.[28]

This reasoning explains also the local minimum in the dependence of the Raman shift on $h_{Ge}$. Since the Ge coverage grows but many of Ge clusters do not contact untill $h_{Ge}$ reaches ~ 8 to 9 Å, stress in a buried Ge layer, and hence, the Raman shift decrease. At $h_{Ge}$ ~ 9 to 10 Å most of Ge QDs start contacting by their bases and many start coalescing. As a result, the stress also begins growing as well as the Raman shift. Finally, at $h_{Ge}$ ~ 10 Å most of Ge clusters coalesce at their bases (Fig. 2i) giving rise to the local maximum of stress in the Ge layer and the local maximum of Raman shift of the Ge–Ge bond vibration band. At higher coverages, Ge clusters goes on coalescing that causes nucleation and growth of extended structural defects that reduces the stress and the Raman shift. In fact, a nanocrystalline film starts to format $h_{Ge}$> 14 Å, which completely forms at $h_{Ge}$~ 18 Å, that obviously further decreases the stress. Besides the nucleation of defects and grain boundaries in Ge layer, tensile-strained Si domains arise in these structures at high Ge coverages at apexes of Ge clusters,[30,33,36] which also contribute to the partial strain relaxation in the Ge layer. Thus, a slightly less blue shift of the Ge–Ge mode in the samples containing thick Ge layers at $h_{Ge}$ = 14 and 18 Å is an expected result related to the additional stress relaxation via the formation of extended defects, grain boundaries and nanocrystals as well as the tensile-strained Si domains (Fig. S3).

The sudden drop of the Ge–Ge peak at $h_{Ge}$ = 14 Å not accompanied by the peak broadening looks like the Ge leakage and needs the further analysis. It is reasonable to suggest that the area under the Ge–Ge peak should increase with the increase of overall Ge content that does not contribute into the Si–Ge signal. The decomposition with the use of two peaks provides a good approximation of the Raman response in the range 290 to 320 cm$^{-1}$ for all the studied samples except for the sample having Ge layers of 18 Å. The peak with the maximum at ~ 302 cm$^{-1}$ is assigned to Si 2TA phonons while the peak at 309–312 cm$^{-1}$ corresponds to compressively stressed Ge. The signal from strained Ge drops significantly on passing from $h_{Ge}$ = 10 Å to $h_{Ge}$ = 14 Å. At the same time, the area under the peak at ~ 302 cm$^{-1}$ increases by 1.7 times and its maximum shifts to ~ 303.5 cm$^{-1}$ (Fig. 6). We suppose that the formation of defects in the structures with high thickness of Ge layers leads to the formation of domains of relaxed Ge. The Raman response from the relaxed Ge overlaps the Si 2TA peak and cannot be well resolved in the spectrum of the sample with $h_{Ge}$ = 14 Å. However, it starts dominating in the sample with $h_{Ge}$ = 18 Å and we have to use three peaks for the decomposition to achieve a good approximation of the experimental data. In this case, the Si 2TA peak shifts back to ~ 302 cm$^{-1}$ and its area decreases to the previous values. The given explanation allows one to avoid the assumption on Ge escapesince the total Ge content not involved into SiGe solution and estimated from the total area under the peaks assigned to both stressed and relaxed Ge becomes proportional to the thickness of deposited Ge.

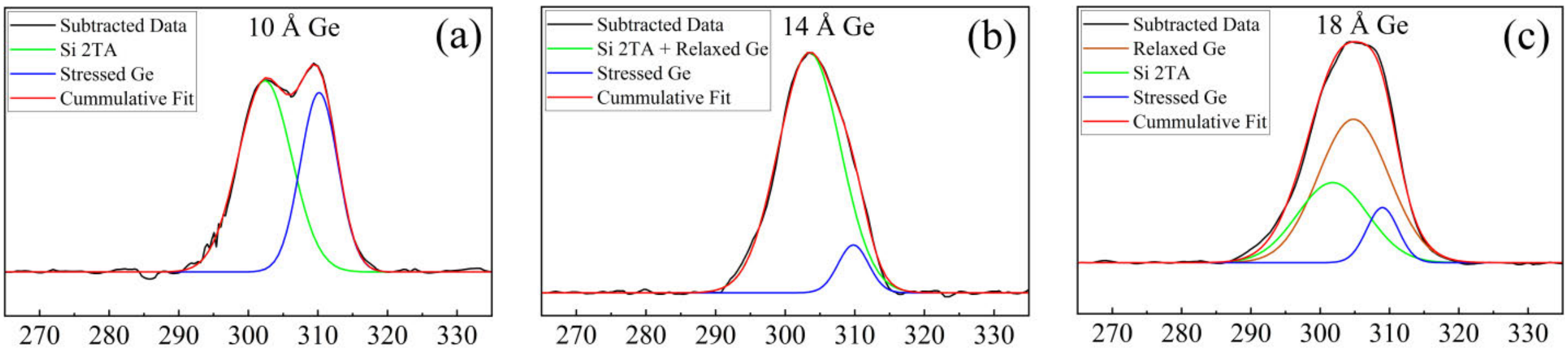

Fitting Results

| Peak | Area Intg | FWHM | Center Grvty |
|---|---|---|---|
| Si 2TA | 943.6740 | 9.0322 | 302.37485 |
| Stressed Ge | 689.5043 | 6.3627 | 310.16511 |

Fitting Results

| Peak | Area Intg | FWHM | Center Grvty |
|---|---|---|---|
| Si 2TA + Relaxed Ge | 1875.781 | 10.518 | 303.47349 |
| Stressed Ge | 154.4559 | 5.1490 | 309.78022 |

Fitting Results

| Peak | Area Intg | FWHM | Center Grvty |
|---|---|---|---|
| Relaxed Ge | 1493.002 | 12.046 | 304.80961 |
| Si 2TA | 847.3678 | 11.982 | 301.80578 |
| Stressed Ge | 246.5714 | 5.3552 | 308.98075 |

**FIGURE 6** The decomposition of Raman response in the range 290 to 320 $cm^{-1}$ by Gaussian peaks for the samples with $h_{Ge}$ = 10 Å (a), 14 Å (b) and 18 Å (c). Fitting results demonstrate integral area, full width at half maximum and the position of maximum for every peak.

## 3.3 | Si–Ge band analysis

The peak related to the Si–Ge mode is hardly observable in samples with $h_{Ge}$ < 10 Å (Fig. 4b). However, one can recognize tiny peaks at ~ 420 $cm^{-1}$ for the samples with Ge coverages of 4, 6 and 7 Å while peaks for the samples with $h_{Ge}$ of 8 and 9 Å are much broader and red shifted as well as that for the sample with $h_{Ge}$= 14 Å (Fig. 4b). We also connect this effect with the cluster number density in arrays, their sizes and hence the distance between QDs.

At low coverages ($h_{Ge}$ = 4 and 6 Å), when QDs are absent or small and the distances between them are large enough (Fig. 2a,b), we deal with Raman scattering by the Si–Ge interfacial bonds and those in thin flat domains of the mixed SiGe layer over the wetting layer provided by the slight intermixing of Ge and Si atoms during the initial stage of coverage by Si.[38–41]

At $h_{Ge}$ = 6 and 7 Å, the band intensity gradually grows compared to that of the sample with $h_{Ge}$ = 4 Å due to the continuous growth of area of the {105} facets of Ge clusters and therefore the increased Ge diffusion from them.[28] However, the Si–Ge vibration band is unshifted that means that the Ge-Si bonds are practically unloaded at these coverages.

Another interesting point arises when comparing the samples with $h_{Ge}$ = 4 and 6 Å considering both Ge–Ge and Si–Ge peaks. The signal from Ge–Ge mode decreases while that from Si–Ge mode increases in passing from 4 to 6 Å. This fact is in good agreement with the suggestion that the diffusion between Si and Ge atoms takes place mostly at {105} facets[28] that leads to enhanced intermixing between Si and Ge when quantum dots start to form. Furthermore, small quantum dots do not contribute to the signal from the Ge–Ge mode and even decrease it because they appear to be nearly totally mixed as well as the wetting layer around them.

At $h_{Ge}$ = 8 and 9 Å, the Si–Ge vibration band is seen to be significantly broadened and red shifted. At these coverages the {105} facets absolutely dominate at the entire Ge/Si interface area (Fig. 2) and hence their contribution to the interfacial stress is crucial. We suggest that the Si–Ge bonds become stretched close to the {105} facets due to stress in a thin interfacial layer between unstrained Ge clusters and unstrained Si, wherein a significant tensile strain was reported in our previous publications.[30,33,36] Moreover, at these coverages, flat regions between quantum dots vanish (Fig. 2) and SiGe layer becomes completely shredded. That leads to the broadening of the Si–Ge peak, as it is seen from the numerical calculations shown below.

The coverage of 10 Å was previously determined to be special since Ge QD arrays formed at 360 °C are the most homogeneous at this value of $h_{Ge}$.[29] In addition, the number density of wedge-like huts reaches maximum and pyramids nearly disappear at $h_{Ge}$ = 10 Å.[29] Here we also see that $h_{Ge}$ = 10 Å is a special point. The Si–Ge Raman band as well as the Ge–Ge one become very intense at this point. Additionally, the Si–Ge band becomes unshifted again and its shape resembles that of the band recorded at $h_{Ge}$ = 7 Å. We suggest that this effect emerges due to the onset of hut coalescing since the wetting layer practically disappears and nearly all huts reach their maximum height at this moment. At this coverage, extended flat domains of Ge and SiGe simultaneously form that provides phonon propagation to quite long distances facing no hindrance (Fig. 7). In the recent work,[28] we directly observed the formation of an extended transition layer of SiGe using a transmission electron microscope, estimated its thickness and determined the minimal thickness of the Si capping layer, at which the enhanced strain-induced diffusion occurs. At $h_{Ge}$ = 10 Å, an array of QDs nearly completely consists of {105} facets, edges and ridges, where from the diffusion of Ge atoms is maximum due to a high stress at the Ge/Si interface.[28] At the same time, according to Ref. 28, an intermediate SiGe layer forms in 50 Å thick Si spacers, which is somewhat stretched compared to Si in the [001] direction, yet keeping the Si lattice period in the [110] directions. The stress-stimulated diffusion of Ge from the {105} facets of huts as well as side and top edges to Si may be responsible for the intensified Raman scattering while the intermediate SiGe layer occurrence might cause the Si–Ge Raman peak returning to the initial position.

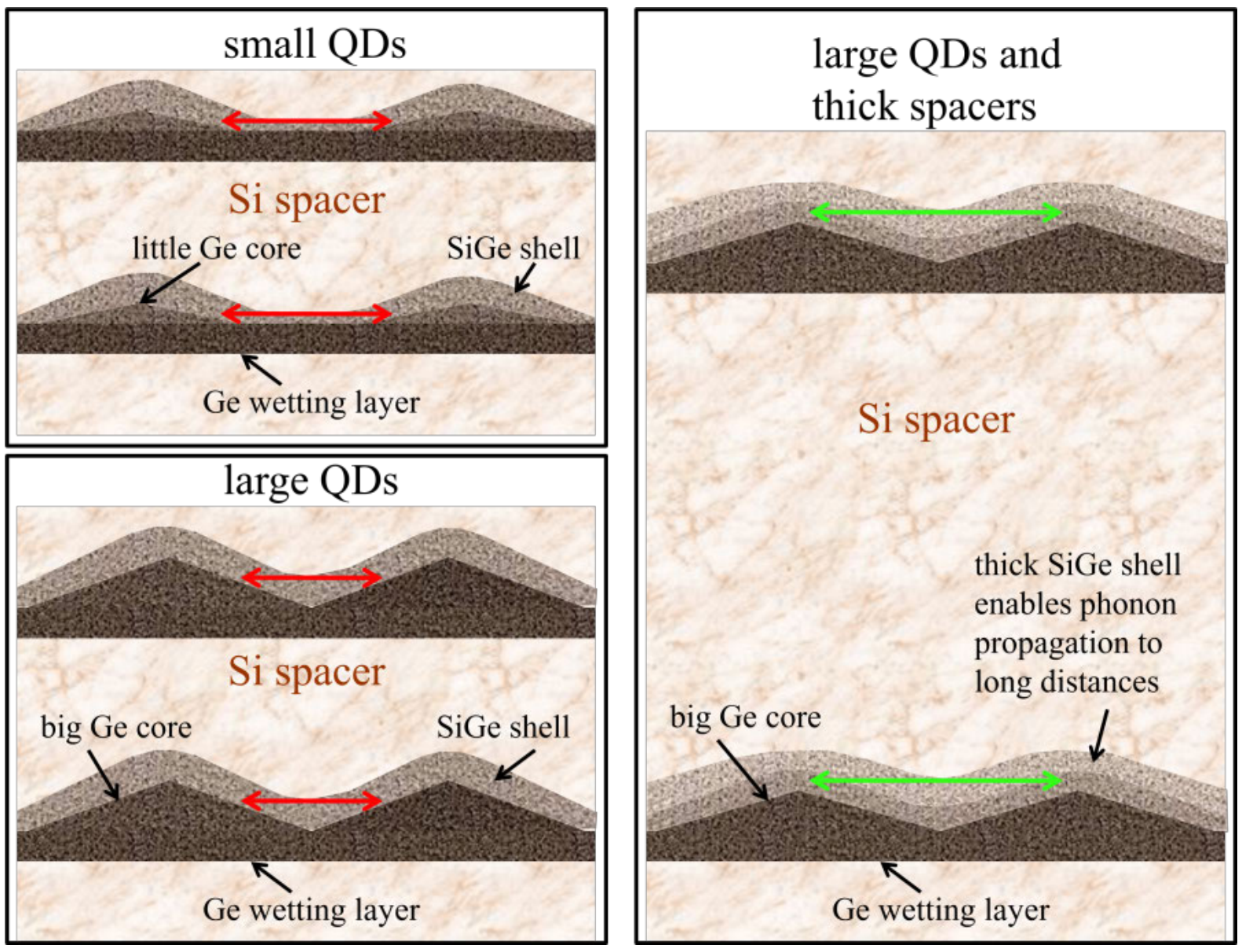

**FIGURE 7** Drawings illustrating the evolution of QDs that explains why the Raman peak of the SiGe mode broadens and weakens with the increasing thickness of deposited Ge until it reaches 10 Å and then suddenly becomes strong and sharp if the thickness of Si spacers is enough to provide strain induced diffusion.

Note that the change of the position and shape of the Si–Ge peak correlates with the shift of the Ge–Ge peak when $h_{Ge}$ rises from 7 to 8 Å. The intensification of the Si–Ge Raman scattering and its peak shift back to the higher wavenumbers correspond with the growth of the Ge–Ge peak Raman shift (Fig. 4a).

For the samples with thick layers of Ge ($h_{Ge}$ = 14 and 18 Å) the Si–Ge peak is also much broader and red shifted in comparison with the sample having layers with $h_{Ge}$ = 10 Å. This may also be explained by a high density of tensile strained domains in Si spacers (Fig. S3a), which make the SiGe layer "shredded" and provide its relaxation. In both these cases, the SiGe layer can be described as an ensemble of nanoparticles. The type stacking fault leading to relaxation in crystal lattice can be recognized in the close-up TEM image (Fig. S3b). Note that defects also present in close proximity to some of QDs in the structures with lower Ge coverage (8 - 10 Å) (Fig. S3c). However, the density of these defects is low and they do not propagate through Si spacers.

### 3.4 | Numerical simulation

Raman peak broadening caused by the optical phonon confinement is an awaited result for Ge/Si structures with Ge nanoclusters.[42–45] However, while QDs are small enough to employ the theory of Raman scattering in nanoparticles they are embedded in a bulk material with very similar crystal lattice and hardly can be treated as isolated nanoparticles. The theoretical approach using the calculation of surface phonon modes[46–48] to considering so thin layers embedded in a bulk material cannot be employed as well. The generation of localized phonon modes is significantly suppressed due to damping[49,50] caused by Si lying along both sides of Ge unlike the case of atoms situated on the surface, which are free in one direction.

We have carried out calculations for model samples, which represent SiGe layers and nanoparticles embedded into Si crystal on the assumption that flat domains of SiGe must exhibit stronger and sharper Raman signal than segmental ones. Atoms in supercells are packed in such a way that both the number of Ge atoms and the number of Si–Ge bonds per supercell are the same for the model samples. Fragments of crystals used in the calculations consisting of several supercells for better demonstration of the atomic structure are depicted in Fig. 8. The first crystal (Fig. 8b) has infinite monoatomic layers of Ge atoms not bounded with each other situated along (100) planes and separated by 7 monoatomic layers of Si while the second (Fig. 8d) has symmetrical inclusions of SiGe separated by Si in all directions.

The Raman spectra derived from calculations via CASTEP method are presented in Fig. 8. The Si–Ge peak broadening in passing from the structure with flat SiGe layers (Fig. 8a) to the structure with segmental SiGe nanoparticles (Fig. 8c) embedded into Si matrix is clearly observable. This proves the suggestion that the shape of the Raman peak from the Si–Ge mode indicates the flatness and homogeneity of SiGe layers present in the structure. We do not compare the calculated positions of Si–Ge peaks with the experimental ones because the overall content of Ge in the model samples is much greater than that in the experiment. It would require a much greater supercell to provide the computation for a comparable Ge content that cannot be done within a reasonable machine time.

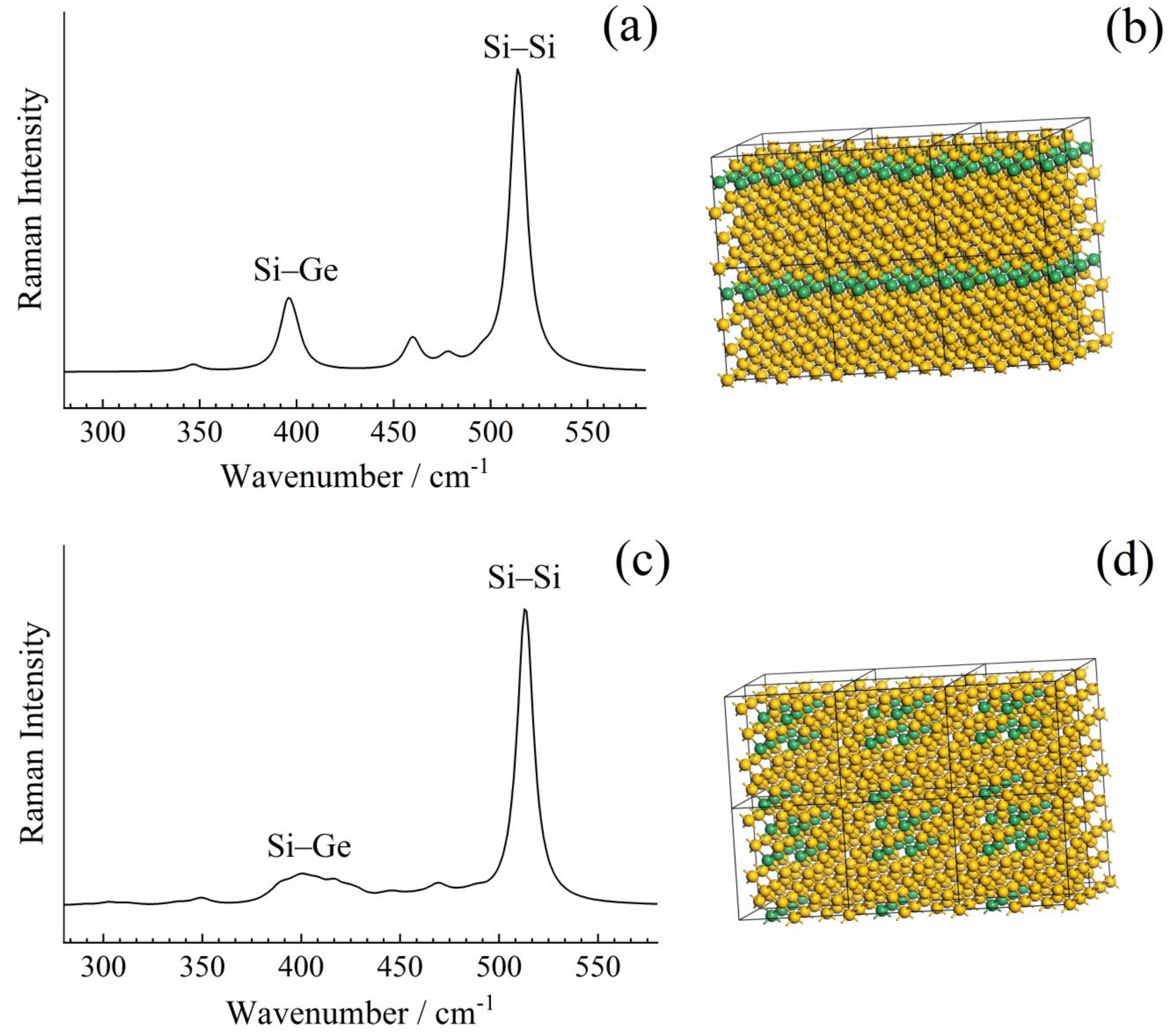

**FIGURE 8** Raman spectra are derived from the calculation of model crystals having flat layers of SiGe (a) or SiGe nanoparticles (c) embedded into Si matrix and corresponding structures of the model crystals used in the numerical simulations; the crystals represent diamond lattice formed by Si (yellow bubbles) and Ge (green bubbles) atoms; Ge atoms are situated along (100) planes (b) or agglomerate into separated SiGe particles (d); in both cases, each Ge atom is bonded with 4 Si atoms.

Moreover, we have made calculations using similar supercells, which had been manually deformed in addition to the standard geometry optimization or contained point defects. Raman spectra of such model structures appear to be dreadfully distorted, therefore we conclude that samples studied in the experiment do not have significant local deformations and point defects of high concentration. This statement does not contradict the experimental results for the structures having Ge layers of 14 and 18 Å in thickness since the average distance between defects observed in these structures is much larger (Fig. S3) than the size of supercells used in the computations.

## 4 | CONCLUSIONS

In summary, we have studied a large number of samples containing thin Ge layers of different thickness deposited on Si(001) and capped by Si. We have discovered some specific features of Raman spectra concerning the Ge–Ge and Si–Ge vibrational modes. Those are discontinuities in the evolution of peak positions of both the Ge–Ge and Si–Ge bands in the interval from 7 to 8 Å of deposited Ge layer effective thickness. These features have been explained in terms of the size and density of self-assembled Ge 3D clusters in the layer. The evolution of the Si–Ge peak shape is controlled by the flatness of the SiGe transitional layer. The latter conclusion is supported by the numerical simulations of Raman scattering that demonstrate the characteristic features arising in the spectra for model structures representing flat and shredded SiGe layers embedded into a Si crystal. A dramatic increase in the intensity of both Ge–Ge and Si–Ge bands for the structure containing Ge layers of 10 Å has been observed and explained in terms of the formation of extended flat layers of SiGe resulted from the coalescence of Ge QDs and the enhanced stress-induced diffusion and intermixing of Si and Ge at {105} facets of Ge huts.

We believe that the reported in this work results can be used not only for Ge/Si heterostructures but also for any crystalline system comprising thin layers strained due to defects or inclusions of a substance with a similar crystal lattice.

## ACKNOLEDGMENTS

S. M. Novikov, V. S. Volkov and A. V. Arsenin gratefully acknowledge the financial support from the Ministry of Science and Higher Education of the Russian Federation (Agreement No.075-15-2021-606) and the Russian Foundation for Basic Research (20-07-00840).

**TABLE S1** Calculated dependence of the Ge–Ge line position on the lateral compression of the Ge layer.

| Effective pressure, GPa | 0 | 0.5 | 1 | 2 | 3 | 4 |
|---|---|---|---|---|---|---|
| Lattice parameter, Å | 5.6575 | 5.6378 | 5.6197 | 5.5847 | 5.5510 | 5.5187 |
| Compression, % | 0 | 0.3 | 0.6 | 1.3 | 1.9 | 2.5 |
| Line position, $cm^{-1}$ | 302.9 | 306.7 | 308.4 | 311.0 | 312.8 | 315.8 |

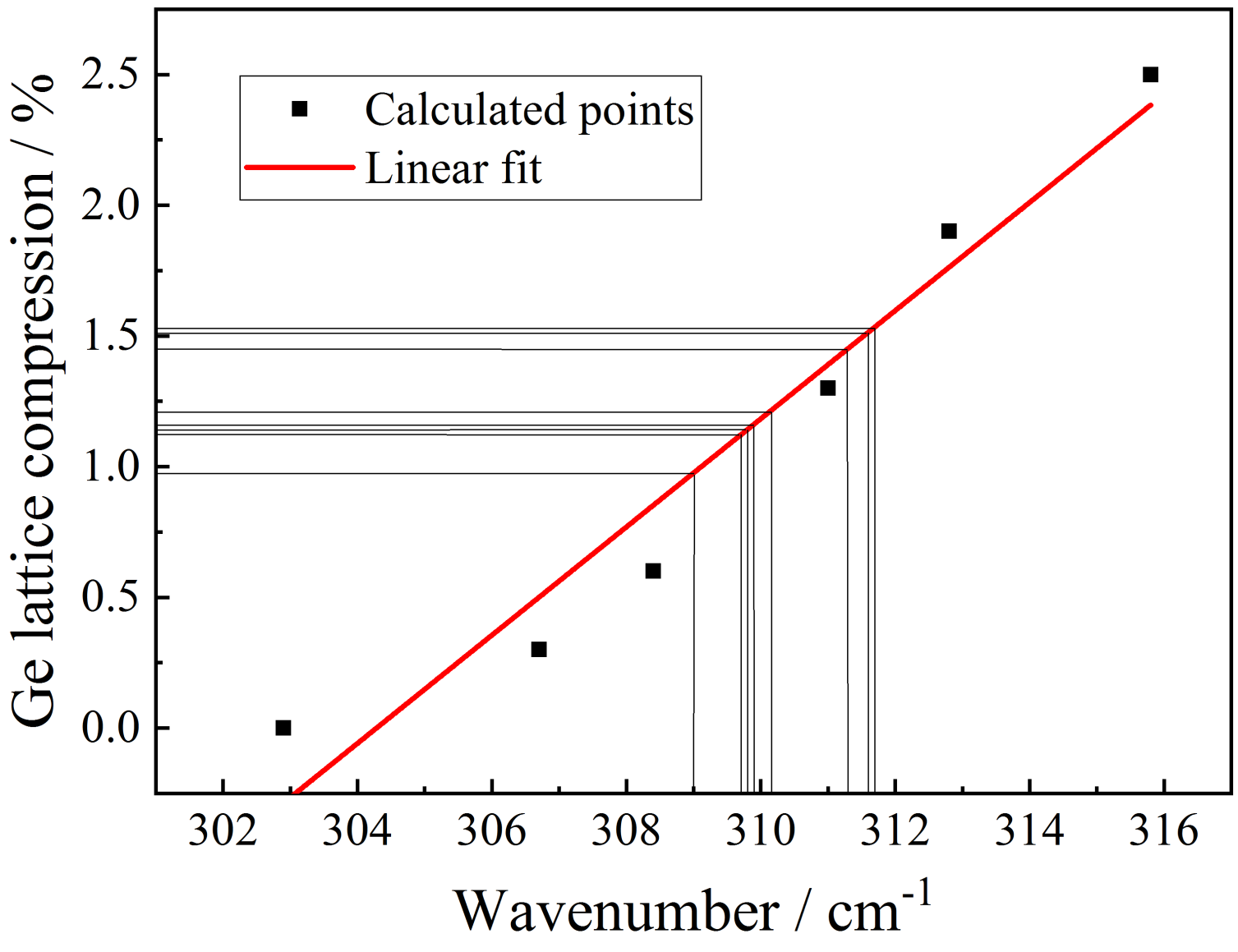

**FIGURE S1** Dependence of Ge lattice compression on Ge–Ge line position derived from numerical simulation; vertical and horizontal lines correspond to the experimental samples.

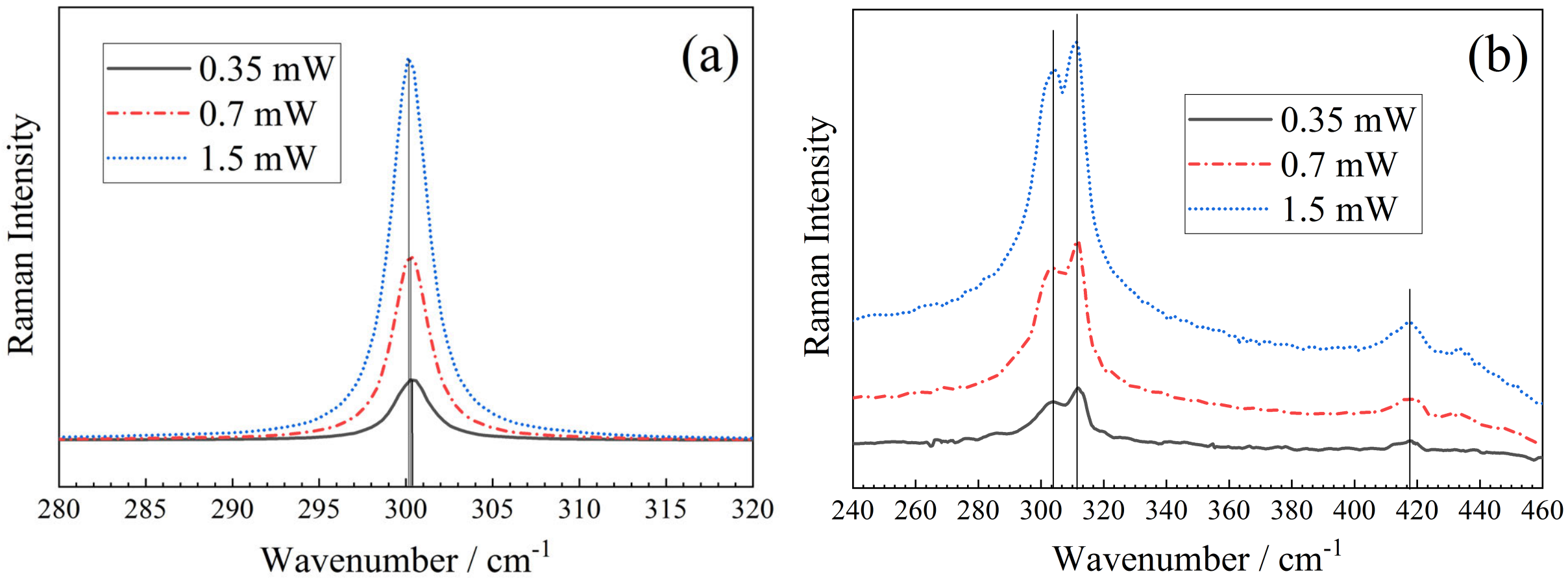

**FIGURE S2** Raman spectra of the Ge reference sample (a) and of the SiGe heterostructure with Ge layers of 10 Å in thickness (b) measured at different incident power.

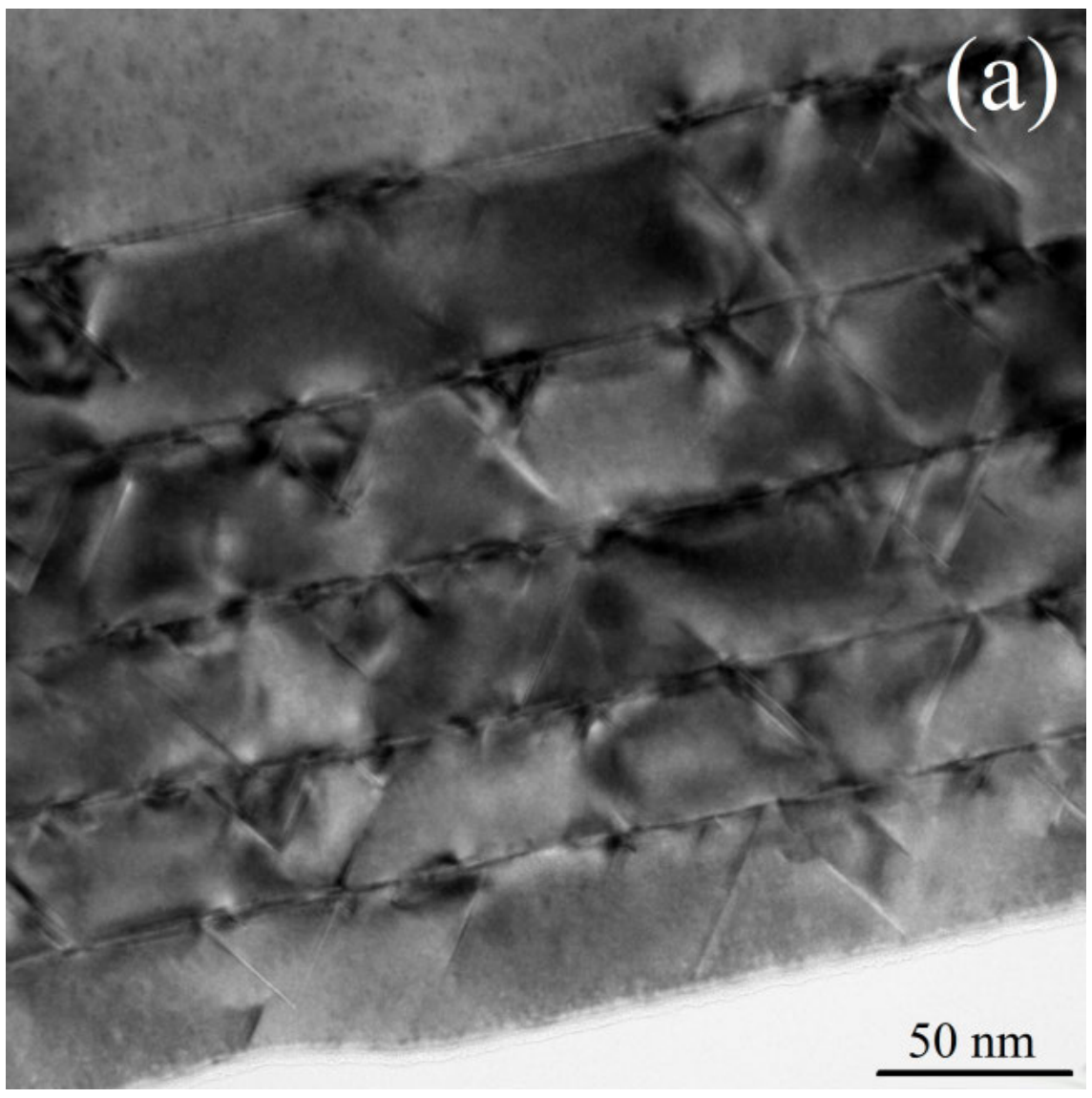

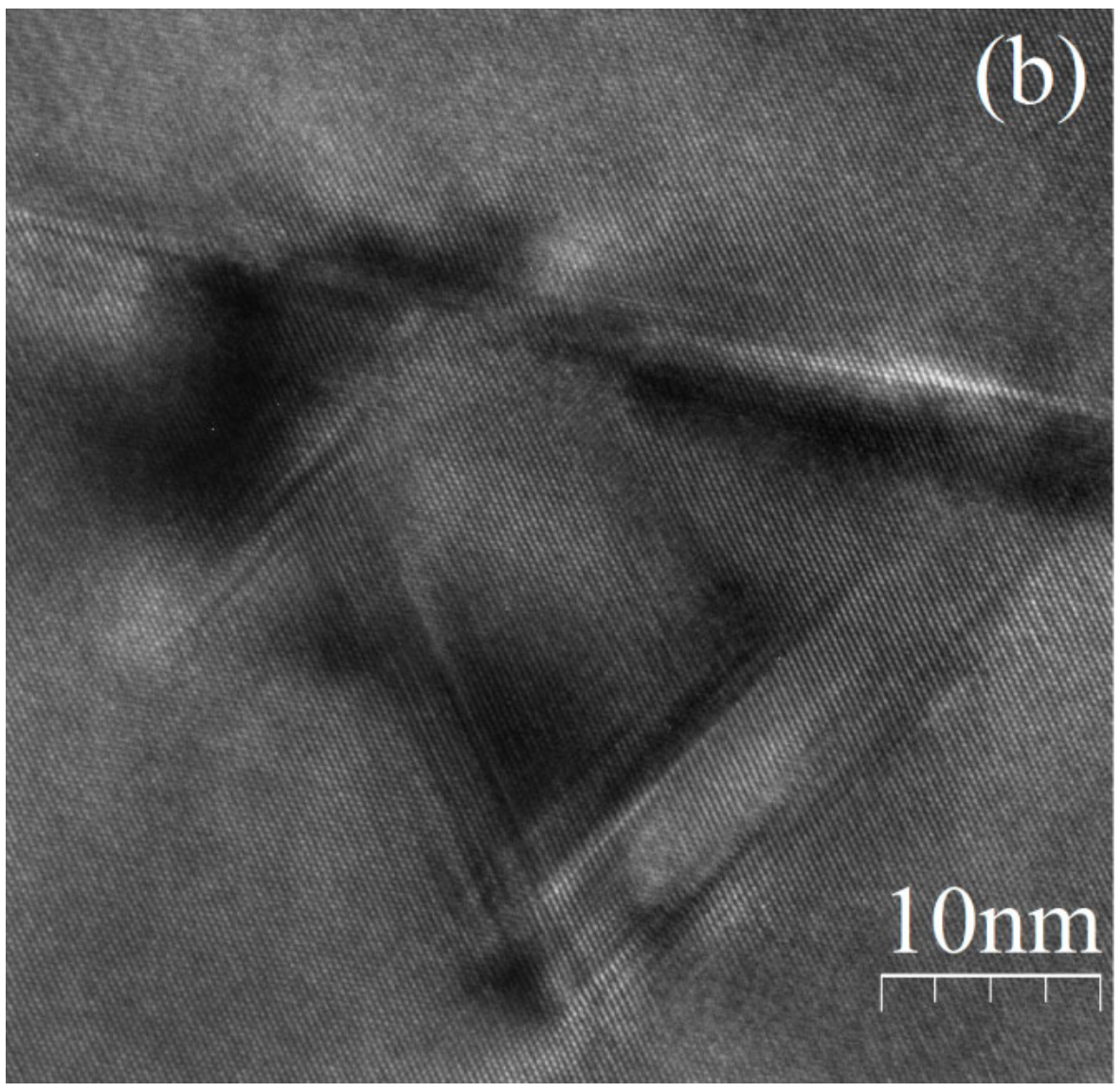

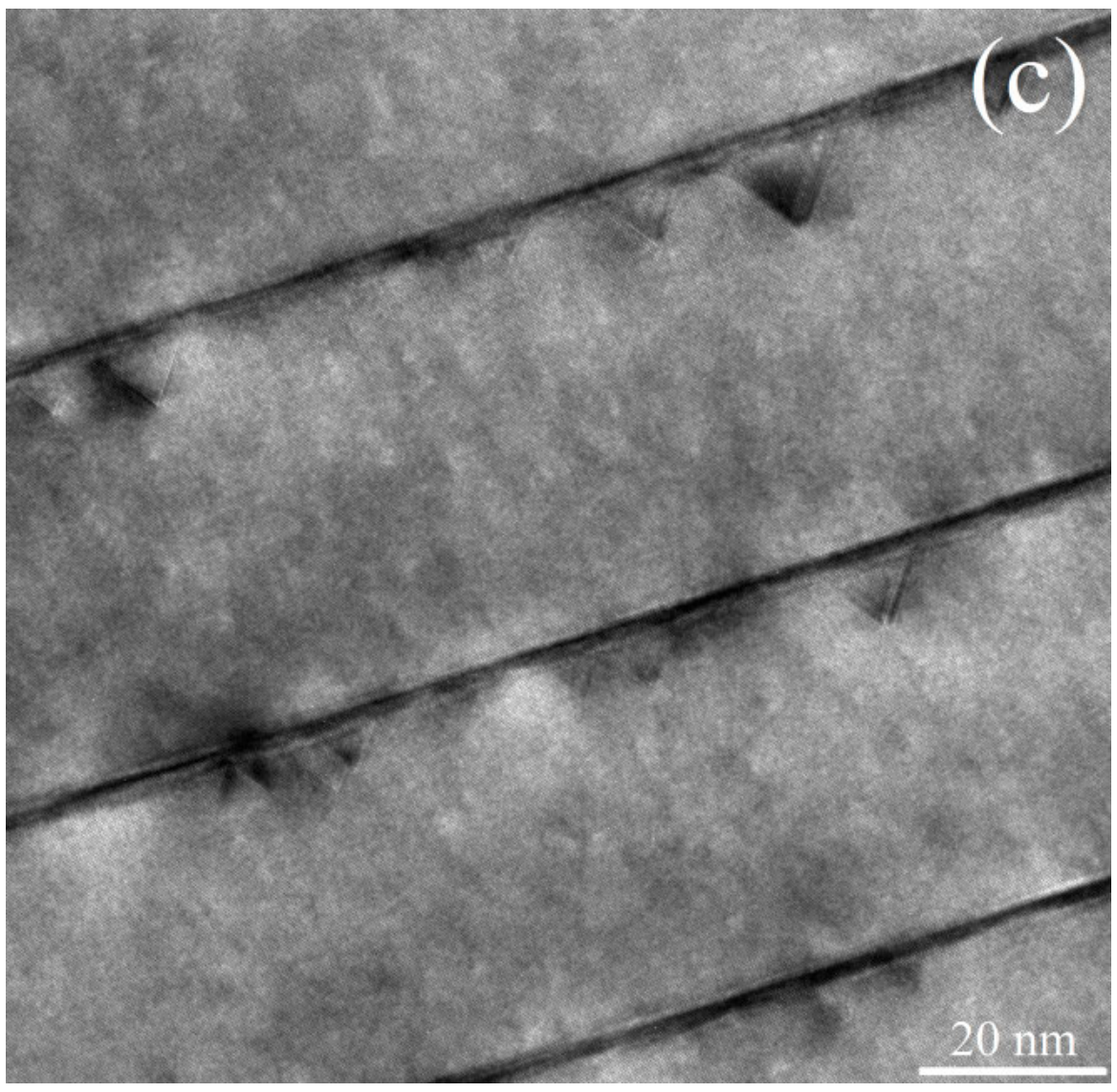

**FIGURE S3** TEM images of the Ge/Si(001) heterostructures containing 5 layers of Ge with $h_{Ge}$ = 18 (a, b) and 9 Å (c).